%% file: genAEPjournalLong.tex
\documentclass[onecolumn,oneside,12pt]{IEEEtran}

\input{mathheader}
\def\be{\begin{eqnarray}}
\def\ee{\end{eqnarray}}
\def\ben{\begin{eqnarray*}}
\def\een{\end{eqnarray*}}

\begin{document}

\title{The Generalized Asymptotic Equipartition Property:
Necessary and Sufficient Conditions}
\author{Matthew~T.~Harrison,~\IEEEmembership{Member,~IEEE}%
\thanks{This work was supported in part by a National Defense Science and Engineering Graduate Fellowship.  The material in this paper is preceded by a technical report \cite{Har:First:2003}.  Preliminary results were presented at 
\cite{HarKon:Maximum:2002}.}%
\thanks{The author is at the Department of Statistics, Carnegie Mellon University, Pittsburgh, PA 15213 (email: harrison@stat.cmu.edu).}}
\maketitle

\begin{abstract}
Suppose a string $X_1^n=(X_1,X_2,\ldots,X_n)$
generated by a memoryless source $(X_n)_{n\geq 1}$ with distribution $P$ is to 
be compressed with distortion no greater than 
$D\geq 0$, using a memoryless random codebook with distribution $Q$. 
The compression performance is  
determined by the ``generalized asymptotic 
equipartition property'' (AEP), which states 
that the probability of finding a $D$-close match 
between $X_1^n$ and any given codeword $Y_1^n$, 
is approximately $2^{-n R(P,Q,D)}$, where the rate function $R(P,Q,D)$
can be expressed as an infimum of relative
entropies. The main purpose here is to remove various restrictive
assumptions on the validity of this result
that have appeared in the recent literature.  
Necessary and sufficient conditions for the generalized AEP are provided in the general setting of abstract alphabets and unbounded distortion measures.  All possible
distortion levels $D\geq 0$ are considered; 
the source $(X_n)_{n\geq 1}$ can be stationary
and ergodic; and the codebook distribution can have memory. Moreover, 
 the behavior
of the matching probability is precisely characterized, even when the generalized AEP is not valid.
Natural characterizations
of the rate function
$R(P,Q,D)$ are established under equally general conditions.
\end{abstract}


\IEEEpeerreviewmaketitle

%
%

\section{Introduction}

Suppose a random string $X_1^n=(X_1,X_2,\ldots,X_n)$
produced by a memoryless source $(X_n)_{n\geq 1}$ with distribution
$P$ on a source alphabet $S$, is to be compressed 
with distortion no more than some $D\geq 0$ with respect
to a single-letter distortion measure 
$\rho(x,y)$.\footnote{Precise rigorous definitions
are given in the following section.}
The basic information-theoretic model for understanding 
the best performance that can be achieved, is the study
of random codebooks. If we generate memoryless random 
strings $Y_1^n=(Y_1,Y_2,\ldots,Y_n)$ according to some 
distribution $Q$ on the reproduction alphabet $T$, 
we would like to know how many such strings are needed 
so that, with high probability, we will be able to find 
at least one codeword $Y_1^n$ that matches the source 
string $X_1^n$ with distortion $D$ or less.
The crucial mathematical problem in answering this
question is the evaluation of the probability that
a given, typical $X_1^n$, will be $D$-close to a 
random $Y_1^n$.  This probability can be expressed as 
\be
\Prob\{Y_1^n\in B_n(X_1^n,D)\,|\,X_1^n\}
=Q^n\bigl(B_n(X_1^n,D)\bigr)
\label{eq:matchingp}
\ee
where $B_n(X_1^n,D)$ denotes the ``distortion ball''
consisting of all reproduction strings that are
within distortion $D$ (or less) from $X_1^n$;
note that the matching probability in (\ref{eq:matchingp})
is itself
a random quantity, as it depends on the source
string $X_1^n$.

The importance of evaluating (\ref{eq:matchingp}) 
was  already identified by Shannon in his classic 
study of rate-distortion theory \cite{shannon:59},
where he showed that, for the best codebook distribution
$Q=Q^*$, we have,
\be
Q^{*n}\bigl(B_n(X_1^n,D)\bigr)\approx 2^{-nR(P,D)}
\label{eq:shannon}
\ee
where $R(P,D)$ is the rate-distortion function
of the source. 

The more general question of evaluating the matching probability 
(\ref{eq:matchingp}) for distributions $Q$ perhaps different 
from the optimal reproduction distribution $Q^*$,
arises naturally in a variety of contexts, including 
problems in pattern-matching, mismatched codebooks, 
Lempel-Ziv compression, combinatorial optimization on 
random strings, and others; see, e.g.,
\cite{ZhaYanWei:Redundancy:1997}%
\cite{LucSzp:Suboptimal:1997}%
\cite{YanKie:Performance:1998}%
\cite{Kon:Implementable:1999}%
\cite{YanZha:Redundancy:1999}%
\cite{DemKon:Asymptotics:1999}%
\cite{yang-zhang:99c}%
\cite{Chi:First:2001}%
\cite{Szp:Average:2001},
and the review and references in \cite{DemKon:Source:2002}.
In this case, Shannon's estimate (\ref{eq:shannon}) is replaced
by the so-called {\em ``generalized asymptotic equipartition
property''} (or generalized AEP), which states that,
\be 
-\frac{1}{n}\log Q^n\bigl(B_n(X_1^n,D)\bigr) \to R(P,Q,D)
\;\;\;\;
\mbox{a.s.}
\label{eq:AEPintro}
\ee
where ``a.s.'' stands for ``almost surely''
and refers to the random string $X_1^n$.
The rate function $R(P,Q,D)$ is defined
in a way that closely resembles the rate-distortion
function definition,
\ben
R(P,Q,D):= \inf_W H(W\|P\times Q)
\een
where $H(\cdot\|\cdot)$ denotes the
relative entropy, and the infimum is
over all (bivariate) probability 
distributions of random variables
$(U,V)$ with values on $S$ and $T$,
respectively, such that $U$ has distribution
$P$ and the expected distortion $E[\rho(U,V)]\leq D$.
(For a broad introduction to the 
generalized AEP, its applications
and refinements, see \cite{DemKon:Source:2002}
and the references therein.)

The study of the rate function $R(P,Q,D)$ 
and its properties is an important step
in understanding the generalized AEP.
In terms of lossy data compression, it is
not hard to see that $R(P,Q,D)$ is equal
to the compression rate achieved by a
(typically mismatched) random codebook
with distribution $Q$. In view of this,
it is not surprising that the rate-distortion
function turns out to be {\em equal} to $R(P,Q^*,D)$,
when the codebook distribution is chosen 
optimally,
$$R(P,D)=\inf_Q  R(P,Q,D)$$
with the infimum being over all
probability distributions $Q$
on the reproduction alphabet $T$.
Another important and useful observation made
by various authors in the recent literature 
is that $R(P,Q,D)$ can alternatively be expressed 
as a convex dual.

Although much is known about the generalized AEP 
and about $R(P,Q,D)$ \cite{DemKon:Source:2002},
all known results are established under certain
restrictive conditions. In most cases the
codebook distribution is required to be memoryless,
and when it is not, it is assumed that the distortion
measure is bounded. Moreover, only distortion
levels in a certain range are considered,
and the case when 
$$
D = \Dmin(P,Q) := \inf \{D : R(P,Q,D) < \infty\}
$$
is always excluded.

The main point of this paper is to remove these constraints,
and to analyze which (if any) are essential for the validity
of the generalized AEP. Our motivation is twofold. On one hand,
unnecessarily stringent conditions make the theoretical picture 
incomplete. On the other, there are applications which naturally
require more general statements.  For example, in the study of
universal lossy compression, where the source distribution
is not known a priori, how can we assume that the distortion
value chosen will be in the appropriate range and will not
coincide with $\Dmin$? (Specific applications                       
   of the results in this paper to central problems in                          
   universal lossy data compression will be developed in                        
   subsequent work.)  Similarly, the usual constraints on 
the distortion measure may fail to hold even for some basic 
distortion measures, like squared error distortion in the 
case of continuous alphabets. And the lack of information
about the generalized AEP at $D=\Dmin$ makes it difficult 
to draw tight correspondences between lossy and lossless
compression, cf.\ \cite{DemKon:Source:2002}.

Thus motivated, we give {\em necessary and sufficient conditions}
for the generalized AEP in (\ref{eq:AEPintro}),
and we precisely characterize the behavior of the 
matching probability in the pathological situations
when the generalized AEP fails. Our results hold
for {\em all} values of $D$, and they cover arbitrary 
abstract alphabets and distortion measures.   
We also allow the source to be stationary and 
ergodic, and the codebook distribution to have
memory. We similarly extend the characterization
of the rate function $R(P,Q,D)$ to the same level
of generality. We show that it can {\em always}
be written as a convex dual, and that a minimizer 
$W$ in the definition of $R(P,Q,D)$ always exists 
(unless, of course, the infimum is taken over the
empty set).  


Sections \ref{s:iid} and \ref{s:genAEP} contain the main results.  
Section \ref{s:memory} contains generalizations to the case when
the codebook distribution has memory.  The bulk of the paper 
is devoted to proofs, which are collected in Section \ref{s:proofs}.   
Our main mathematical tool is a generalized, one-sided version of 
the G\"artner-Ellis theorem from large deviations.
It is stated and proved in Section~\ref{s:GE}, and it may be of 
independent interest. 
Finally, the important special case when $D=\Dmin$ is analyzed
using results about the recurrence properties of random 
walks with stationary increments.   

%
%

\section{Characterization of the rate function} \label{s:iid}

Let $S$ be the source alphabet with its associated $\sigma$-algebra
$\mathcal{S}$, let $(T,\mathcal{T})$ be the reproduction alphabet,
and take $\rho:S\times T\mapsto [0,\infty)$ to be a distortion measure.  
We only assume that $(S,\mathcal{S})$ and $(T,\mathcal{T})$ are Borel 
spaces\footnote{Borel spaces include ${\mathbb R}^d$ as well as a large
class of infinite-dimensional spaces, including Polish spaces.
This assumption is made so that we can avoid certain 
pathologies while working with random sequences and conditional 
distributions \cite{Kal:Foundations:2002}.} and that $\rho$ is 
$\sigma(\mathcal{S}\times\mathcal{T})$-measurable.  
Henceforth, these $\sigma$-algebras and the various product 
$\sigma$-algebras derived from them are understood from the context.  
We use the abbreviations r.v., a.s., i.o., l.sc., u.sc.~and $\log$ 
for random variable, almost surely, infinitely often, lower semicontinuous, 
upper semicontinuous and $\log_e$, respectively.  
If $U$ and $V$ are r.v.'s and $g(u) := Ef(u,V)$, we use 
the notation $E_Vf(U,V)$ for the r.v.~$g(U)$.   
When $U$ and $V$ are independent, then 
$E_Vf(U,V)\relas{=} E[f(U,V)|U]$.   

We write $X$ and $Y$ for two independent r.v.'s taking values in 
$S$ and $T$, respectively, with $X\sim P$ and $Y\sim Q$.  
We use $\rho$ to define a sequence of single-letter 
distortion measures
$\rho_n$ on $S^n\times T^n$, $n\geq 1$, by
\[ \rho_n(x_1^n,y_1^n) := \frac{1}{n}\sum_{k=1}^n \rho(x_k,y_k)  \]
where $x_i^j:=(x_i,\dotsc,x_j)$.  
The dependence on $\rho$ or $\rho_n$ is suppressed in nearly all 
of our notation.  We use
\[ B_n(x_1^n,D) := \left\{y_1^n\in T^n:\rho_n(x_1^n,y_1^n) \leq D\right\} \]
to denote the distortion ball of radius $D$ around $x_1^n$.

If $W$ is a probability distribution on $S\times T$, then we use $W_S$ to denote the marginal distribution of $W$ on $S$, and similarly for $W_T$.  An important subset of probability distributions on $S\times T$ is
\[ W(P,D) := \left\{W : W_S=P, \ E_{(U,V)\sim W}\rho(U,V) \leq D\right\} . \]
This subset comes up in the definition of the 
rate-distortion function
\[ R(P,D) := \inf_{W\in W(P,D)} H(W\|\matsize{W_S}{W_T})  \]
which we take to be $+\infty$ when $W(P,D)$ is empty.
$H(\mu\|\nu)$ denotes the relative entropy (in nats). 
\[ H(\mu\|\nu) := \begin{cases} E_\mu \log \frac{d\mu}{d\nu} & \text{if $\mu\ll\nu$,} \\ \infty & \text{otherwise}. \end{cases} \]
Note that $H(W\|\matsize{W_S}{W_T})$ is 
the mutual information between r.v.'s $(U,V)$ with joint distribution $W$.
 
Since $H(W\|\matsize{W_S}{W_T})=\inf_Q H(W\|\matsize{W_S}{Q})$, analysis of $R(P,D)$ often proceeds by expanding the infimum into two parts, namely,
\begin{gather*} R(P,D) = \inf_Q R(P,Q,D) \\ R(P,Q,D) := \inf_{W\in W(P,D)} H(W\|\matsize{P}{Q}) . \end{gather*}
The first infimum is over all probability distributions $Q$ on $T$.
Expanding the definition in this way is convenient, because $R(P,Q,D)$ can be expressed as a simple Fenchel-Legendre transform.  In particular, define
\begin{gather*} \Lambda(P,Q,\lambda) := E_{X}\left[\log E_{Y} e^{\lambda\rho(X,Y)}\right] \\  \Lambda^*(P,Q,D) := \sup_{\lambda\leq 0} \left[\lambda D - \Lambda(P,Q,\lambda)\right] . \end{gather*}

\vskip 2ex
\begin{prop} \label{p:RPQD} $R(P,Q,D) = \Lambda^*(P,Q,D)$ for all $D$.  If $W(P,D)$ is not empty, then this set contains a $W$ such that $R(P,Q,D) = H(W\|\matsize{P}{Q})$. \end{prop}
\vskip 2ex

This alternative characterization is well known (see \cite{DemKon:Source:2002} for a review and references).  We state it as a proposition and prove it below because typically it is qualified by other assumptions on $\rho$ and $D$.  In particular, the case $D = \Dmin(P,Q)$ is almost always excluded, where
\[ \Dmin(P,Q) := \inf \{D : R(P,Q,D) < \infty\} . \]

$R(P,Q,D)$ has two other important characterizations that arise in a variety of contexts.  
Let $P_{x_1^n}$ denote the empirical distribution on $S$ of $x_1^n$, let $Q^n$ denote the $n$-times product measure of $Q$ on $T^n$ and define
\[ L_n(x_1^n,Q_n,D) := -\frac{1}{n}\log Q_n\bigl(B_n(x_1^n,D)\bigr) \]
for any probability distribution $Q_n$ on $T^n$. 

\vskip 2ex
\begin{thm} \label{t:genAEP} If $(X_n)_{n\geq 1}$ is stationary and ergodic, taking values in $S$, with $X_1\sim P$, then
\[ \liminf_{n\to\infty} L_n(X_1^n,Q^n,D) \relas{=} R(P,Q,D) \]
for all $D$.  The result also holds with $L_n(X_1^n,Q^n,D)$ replaced by $R(P_{X_1^n},Q,D)$.  
\end{thm}
\vskip 2ex

Of course, if the limit exists, then the $\liminf$ is the also the 
limit and Theorem \ref{t:genAEP} is what Dembo and Kontoyiannis 
\cite{DemKon:Source:2002} call the {\em generalized AEP}.  There are, 
however, pathological situations where the limit does not exist.  
In the next section 
we give necessary and sufficient conditions for the existence of the 
limit and we analyze in detail the situation where the limit does not exist.  

%
%

\section{The Generalized AEP} \label{s:genAEP}

Here and in the remainder of the paper we will always assume that $(X_n)_{n\geq 1}$ is stationary and ergodic, taking values in $S$, with $X_1\sim P$.  
Define\footnote{ The essential infimum of a random variable $\eta$, is $\essinf \eta := \inf \{ r: \Prob\{\eta < r\} > 0\}$.}
\[ \rho_Q(x) := \essinf \rho(x,Y) . \]
We can exactly characterize when the $\liminf$ is actually a limit in Theorem \ref{t:genAEP}.  

\vskip 2ex
\begin{thm} \label{t:iff} $\lim_n L_n(X_1^n,Q^n,D)$ does not exist with positive probability if and only if $0 < D = \Dmin(P,Q) < \infty$ and $R(P,Q,D) < \infty$ and $\rho_Q(X_1)$ is not a.s.~constant.  Furthermore, in this situation
\begin{subequations} \label{e:genAEP_Dmin}
\begin{gather}
\Prob\{L_n(X_1^n,Q^n,D) = \infty \text{ i.o.}\} > 0  \label{e:genAEP_Dmin_infty} \\
\Prob\{L_n(X_1^n,Q^n,D) < \infty \text{ i.o.}\} = 1  \label{e:genAEP_Dmin_finite} \\
\lim_{m\to\infty} L_{N_m}(X_1^{N_m},Q^{N_m},D) \relas{=} R(P,Q,D)  \label{e:genAEP_Dmin_lim}
\end{gather}
\end{subequations}
where $(N_m)_{m\geq 1}$ is the (a.s.)~infinite random subsequence of $(n)_{n\geq 1}$ for which $L_n(X_1^n,Q^n,D)$ is finite.  All of the above also 
holds with $L_n(X_1^n,Q^n,D)$ replaced by $R(P_{X_1^n},Q,D)$.
\end{thm}
\vskip 2ex

Combined with Theorem \ref{t:genAEP}, this gives necessary and sufficient 
conditions for the generalized AEP.  Both theorems are proven below.  
The proof shows that $(N_m)_{m\geq 1}$ can also be (a.s.)~characterized as 
the random subsequence for which
\begin{equation} \frac{1}{n}\sum_{k=1}^n \rho_Q(X_k) \leq D . \label{e:genAEP_essinf} \end{equation}
Note that $\Dmin(P,Q) = E\rho_Q(X_1)$, whenever the former is finite.  

A simple example that illustrates the pathology is the following: Let $(X_n)_{n\geq 1}$ be the sequence $1,0,1,0,\dotsc$ with probability $1/2$ and the sequence $0,1,0,1,\dotsc$ with probability $1/2$, namely, the binary, stationary, periodic Markov chain (which is ergodic).  Let $Q$ be the point mass at $0$, let $\rho(x,y) := |x-y|$ and let $D=1/2$.  Note that $\rho_Q(X_1)=X_1$ is not constant, that $D=\Dmin(P,Q)=1/2$ and that $R(P,Q,D)=0$ is finite.  In the case when $X_1=0$, $L_n(X_1^n,Q^n,D) = 0$ for all $n$.  In the case when $X_1=1$, however, $L_{2n}(X_1^{2n},Q^{2n},D)=0$ and $L_{2n-1}(X_1^{2n-1},Q^{2n-1},D)=\infty$ for all $n$.

%
%

\section{Extensions to the case with memory} \label{s:memory}

Although the source $(X_n)_{n\geq 1}$ can have memory, the 
generalized AEP stated thus far is restricted to the case where the reproduction distribution is memoryless, that is, $L_n$ is evaluated with a product measure $Q^n$.  We relax this assumption here.   

Let $\mathbb{P}$ denote the distribution of $(X_n)_{n\geq 1}$, which we continue to assume is stationary and ergodic with $X_1\sim P$.  Let $\mathbb{Q}$ denote the distribution of a stationary random process $(Y_n)_{n\geq 1}$ taking values in $T$ with $Y_1\sim Q$.  We use $P_n$ and $Q_n$ to denote the distributions of $X_1^n$ and $Y_1^n$, respectively, which are assumed to be independent.  The results stated so far assume that $\mathbb{Q}$ is memoryless, that is, $Q_n=Q^n$.

For the results in this section, however, we assume that $\mathbb{Q}$ satisfies the following strong mixing condition:
\[ C^{-1}\mathbb{Q}(A)\mathbb{Q}(B) \leq \mathbb{Q}(A\cap B) \leq C \mathbb{Q}(A)\mathbb{Q}(B) \]
for some fixed $1 \leq C < \infty$ and any $A\in \sigma(Y_1^n)$ and $B\in \sigma(Y_{n+1}^\infty)$ and any $n$.  Notice that this implies ergodicity and includes the cases where $\mathbb{Q}$ is memoryless~($C=1$) and where $\mathbb{Q}$ is a hidden Markov model (HMM) whose underlying Markov chain has a finite state space with all (strictly) positive transition probabilities.  For the special case of a finite state Markov chain, a formula for $R_\infty(\mathbb{P},\mathbb{Q},D)$ not involving limits was identified in \cite{YanKie:Performance:1998}.

Following the definition of $R(P,Q,D)$, define
\[ R_n(P_n,Q_n,D) := \frac{1}{n} \inf_{W_n\in W_n(P_n,D)} H(W_n\|\matsize{P_n}{Q_n})  \]
where $W_n(P_n,D)$ is the subset of probability distributions on $S^n\times T^n$ defined analogously to $W(P,D)$ except with $\rho_n$ instead of $\rho$.  Also, let $\delta_{x_1^n}$ be the probability distribution on $S^n$ that assigns probability one to the sequence $x_1^n$.

\vskip 2ex
\begin{thm} \label{t:mix}
Theorems \ref{t:genAEP} and \ref{t:iff} remain valid when $Q^n$ is replaced by $Q_n$, $R(P_{X_1^n},Q,D)$ is replaced by $R_n(\delta_{X_1^n},Q_n,D)$ and $R(P,Q,D)$ is replaced by $R_\infty(\mathbb{P},\mathbb{Q},D)$, where
\[ R_\infty(\mathbb{P},\mathbb{Q},D) := \lim_{n\to\infty} R_n(P_n,Q_n,D) . \]
\end{thm}
\vskip 2ex

The existence of the limit in the definition of $R_\infty(\mathbb{P},\mathbb{Q},D)$ is part of the result.  
Define
\[ \Dmin(\mathbb{P},\mathbb{Q}) := \inf\{ D: R_\infty(\mathbb{P},\mathbb{Q},D) < \infty\} . \]
Note that the mixing conditions here are strong enough to ensure that
\begin{equation} \label{e:DminDmin}
\Dmin(P,Q) = \Dmin(\mathbb{P},\mathbb{Q})
\end{equation}
and that
\begin{equation} \label{e:essinf}
\essinf \rho_n(x_1^n,Y_1^n) = \frac{1}{n}\sum_{k=1}^n \rho_Q(x_k)
\end{equation}
which is why the results for memory can still be in terms of $\Dmin(P,Q)$ and $\rho_Q$.  Extending Theorem \ref{t:iff} to situations where these do not hold seems difficult.  The generalized AEP for $\mathbb{Q}$ with memory can also be found in \cite{Chi:First:2001,Chi:Stochastic:2001,DemKon:Source:2002} under more general mixing conditions but for bounded distortion measure $\rho$ and for $D\neq\Dmin(\mathbb{P},\mathbb{Q})$.  

Define
\begin{gather*} \Lambda_n(P_n,Q_n,\lambda) := E_{X_1^n}\left[\log E_{Y_1^n} e^{\lambda \rho_n(X_1^n,Y_1^n)}\right] \\ \Lambda_n^*(P_n,Q_n,D) := \frac{1}{n}\sup_{\lambda\leq 0} \left[\lambda D - \Lambda_n(P_n,Q_n,\lambda)\right] . \end{gather*}
Proposition \ref{p:RPQD} immediately gives \[ R_n(P_n,Q_n,D) = \Lambda_n^*(P_n,Q_n,D) \] so $R_\infty(\mathbb{P},\mathbb{Q},D)$ is the limit of a sequence of Fenchel-Legendre transforms.  Analogous to the memoryless case, it can also be characterized directly as a Fenchel-Legendre transform.  

\vskip 2ex
\begin{prop} \label{p:Rmix} Define
\begin{gather*} \Lambda_\infty(\mathbb{P},\mathbb{Q},\lambda) := \lim_{n\to\infty} \frac{1}{n}\Lambda_n(P_n,Q_n,n\lambda) \\ \Lambda_\infty^*(\mathbb{P},\mathbb{Q},D) := \sup_{\lambda\leq 0} \left[\lambda D - \Lambda_\infty(\mathbb{P},\mathbb{Q},\lambda)\right] . \end{gather*}
Then $R_\infty(\mathbb{P},\mathbb{Q},D) = \Lambda_\infty^*(\mathbb{P},\mathbb{Q},D)$.
\end{prop}
\vskip 2ex

The existence of the limit in the definition of $\Lambda_\infty(\mathbb{P},\mathbb{Q},\lambda)$ is part of the result.  Occasionally it is more convenient to rewrite
\begin{equation} \label{e:LL*} \Lambda_n^*(P_n,Q_n,D) = \sup_{\lambda\leq 0} \left[\lambda D - \frac{1}{n}\Lambda_n(P_n,Q_n,n\lambda)\right] . \end{equation}
This form makes it easy to show that $R_n(P_n,Q^n,D)=R(P,Q,D)$ and that $R_n(\delta_{x_1^n},Q^n,D)=R(P_{x_1^n},Q,D)$, so that whenever $\mathbb{Q}$ is memoryless, $R_\infty(\mathbb{P},\mathbb{Q},D)=R(P,Q,D)$ and all the results coincide.  

%
%

\section{Proofs} \label{s:proofs}

The proofs occasionally refer to $\Dave(P,Q) := E\rho(X,Y)$ for independent $X\sim P$ and $Y\sim Q$.   
\subsection{Properties of $\Lambda$ and $\Lambda^*$ for arbitrary distortion measures} \label{s:Lambda}

A common assumption in the literature is that $\rho$ is either bounded or satisfies some moment conditions, such as $\Dave(P,Q)<\infty$.  Since we do not assume these things here, we need to reverify many properties of $\Lambda$ and $\Lambda^*$ that can be found elsewhere under stronger conditions.  These properties lead to the generalized AEP under the usual condition that $D\neq\Dmin$.  More detailed proofs, including measurability issues, can be found in a technical report that preceded this paper \cite{Har:First:2003}.  

In this section we will use the assumptions and notation from Section \ref{s:iid}, however, we will suppress the dependence on $P$ and $Q$ whenever possible.  In particular, we will think about $\Lambda(\lambda) := \Lambda(P,Q,\lambda)$ and $\Lambda^*(D) := \Lambda^*(P,Q,D)$ as functions of $\lambda$ and $D$, respectively.  It is also convenient to temporarily redefine
\[ \Dmin := \inf\{D: \Lambda^*(D) < \infty\} \]
until the end of this section where we prove Proposition \ref{p:RPQD}.  Proposition \ref{p:RPQD} shows that $\Lambda^*(D)=R(P,Q,D)$, so both definitions of $\Dmin$ are equivalent.  Note that everything in this section applies equally well to $\Lambda_n$, $\Lambda_n^*$ and $R_n$ as defined in Section \ref{s:memory}.

We begin with the following Lemma which comes mostly from \cite{DemZei:Large:1998}[Lem.~2.2.5, Ex.~2.2.24].  See also \cite{YanZha:Redundancy:1999,DemKon:Source:2002}.  

\vskip 2ex
\begin{lem} \cite{DemZei:Large:1998} \label{l:derivative}
Let $Z$ be a real-valued, nonnegative random variable.  Define
\[ \Gamma(\lambda) := \log Ee^{\lambda Z} . \]
$\Gamma$ is nondecreasing and convex.  $\Gamma$ is finite, nonpositive and $C^\infty$ on $(-\infty,0)$ with
\[ \lim_{\lambda\uparrow 0}\Gamma(\lambda) = \Gamma(0) = 0 \quad \text{and } \quad \Gamma'(\lambda) = \frac{EZe^{\lambda Z}}{Ee^{\lambda Z}} , \quad \lambda < 0 . \]
$\Gamma'$ is finite, nonnegative and nondecreasing on $(-\infty,0)$ with 
\[ \lim_{\lambda\downarrow -\infty} \Gamma'(\lambda) = \essinf Z \quad \text{ and } \quad \lim_{\lambda\uparrow 0} \Gamma'(\lambda) = EZ . \]  If $\essinf Z < EZ$, then $\Gamma$ is strictly convex on $(-\infty,0)$.
\end{lem}
\vskip 2ex

Define $\Gamma(\lambda,x) := \log Ee^{\lambda\rho(x,Y)}$.  For fixed $x$, we can apply Lemma \ref{l:derivative} to the r.v.~$Z:=\rho(x,Y)$ to get several regularity properties of $\Gamma(\cdot,x)$.  It turns out that these regularity properties are preserved by expectations, i.e., they continue to hold for $\Lambda(\lambda)=E\Gamma(\lambda,X)$.  A sufficient condition is that $\Lambda$ be finite on $(-\infty,0]$.  This replaces the typical moment conditions on $\rho$.  Note that if $\Lambda^*(D)$ is finite for some $D$, i.e., if $\Dmin$ is finite, then this condition is trivially satisfied.  

\vskip 2ex
\begin{lem} \label{l:Lambda}
$\Lambda$ is nondecreasing and convex.  Suppose $\Lambda$ is finite on $(-\infty,0]$.  Then $\Lambda$ is nonpositive and $C^1$ on $(-\infty,0)$ with $\lim_{\lambda\uparrow 0}\Lambda(\lambda) = \Lambda(0) = 0$ and \[ \Lambda'(\lambda) = E_X\left[\frac{E_Y\rho(X,Y)e^{\lambda \rho(X,Y)}}{E_Ye^{\lambda \rho(X,Y)}}\right] , \quad \lambda < 0 . \]
$\Lambda'$ is finite, nonnegative and nondecreasing on $(-\infty,0)$ with 
\[ \lim_{\lambda\downarrow -\infty} \Lambda'(\lambda) = E\rho_Q(X) \quad \text{ and } \quad \lim_{\lambda\uparrow 0} \Lambda'(\lambda) = \Dave . \]  If $E\rho_Q(X) < \Dave$, then $\Lambda$ is strictly convex on $(-\infty,0)$.  
\end{lem}
\vskip 2ex

\begin{IEEEproof}
The statements about $\Lambda$ are trivial.  We will focus on the properties of $\Lambda'$ which follow more or less immediately from the convexity of $\Lambda$ and the differentiability of $\Gamma(\cdot,x)$.  Let $\Lambda'_-$ and $\Lambda'_+$ be the left hand and right hand derivatives of $\Lambda$, respectively, which are finite for $\lambda < 0$.  The monotone convergence theorem immediately gives $\Lambda'_-(\lambda)=E\Gamma'(\lambda,X)$ for $\lambda<0$.  (The same argument can be used as $\lambda\uparrow 0$.)  This shows that $\Gamma'(\lambda,X)$ has finite expectation and lets us use the dominated convergence theorem to get that $\Lambda'_+(\lambda) = E\Gamma'(\lambda,X)$.  (The same argument can be used as $\lambda\downarrow -\infty$.) So the left and right hand derivatives of $\Lambda$ are identical and have the given form.  Recall that a differentiable, convex function has a continuous derivative.
\end{IEEEproof}

These properties of $\Lambda$ give the following well known properties of $\Lambda^*$, which we state without proof, except for \eqref{e:Dmin}.  See \cite{DemZei:Large:1998}[Lem.~2.2.5] and \cite{Roc:Convex:1970}[Thm.~23.5, Cor.~23.5.1, Thm.~25.1].  

\vskip 2ex
\begin{lem} \label{l:Lambda*} $\Lambda^*$ is convex, l.sc., nonnegative, nonincreasing and continuous from the right.  $\Lambda^*\equiv \infty$ on $(-\infty,\Dmin)$ and $\Lambda^*\equiv 0$ on $[\Dave,\infty)$.  If $D\leq \Dave$, then $\Lambda^*(D)=\sup_{\lambda\in\mathbb{R}}[\lambda D-\Lambda(\lambda)]$.  If $\Dmin<\infty$ (so that Lemma \ref{l:Lambda} applies), then $\Dmin=E\rho_Q(X)$, $\Lambda^*$ is finite and $C^1$ on $(\Dmin,\infty)$ and
\begin{equation} \label{e:Dmin}
\Lambda^*(\Dmin) = E_X\left[-\log E_Y\ind\{\rho(X,Y)=\rho_Q(X)\}\right] .
\end{equation}  If further $\Dmin < \Dave$, then $\Lambda^*$ is strictly convex (and thus strictly decreasing) on $(\Dmin,\Dave)$ and for each $D\in(\Dmin,\Dave)$ there exists a unique $\lambda_D<0$ such that $\Lambda^*(D) = \lambda_D D-\Lambda(\lambda_D)$.
\end{lem}
\vskip 2ex

\begin{IEEEproof}  We only prove \eqref{e:Dmin}.
Define  \[ \tilde\rho(x,y) := \max\{\rho(x,y)-\rho_Q(x),0\} \] so that $\tilde\rho$ is a valid distortion measure and so that \[ \rho(x,Y)\relas{=}\tilde\rho(x,Y)+\rho_Q(x) . \]  Let $\tilde\Lambda$ be defined analogously to $\Lambda$, except with $\tilde\rho$ instead of $\rho$.  We have $\Lambda(\lambda) = \tilde\Lambda(\lambda)+\lambda \Dmin$ so that
\begin{align*} & \Lambda^*(\Dmin) = \sup_{\lambda\leq 0} \left[\lambda \Dmin - \tilde\Lambda(\lambda) - \lambda\Dmin\right] = \lim_{\lambda\downarrow -\infty} -\tilde\Lambda(\lambda) \\
& \quad = \lim_{\lambda\downarrow -\infty} E_X\left[-\log E_Ye^{\lambda\tilde\rho(X,Y)}\right] \\
& \quad = E_X\left[-\log E_Y\left(\lim_{\lambda\downarrow -\infty} e^{\lambda\tilde\rho(X,Y)}\right)\right] \\
& \quad = E_X\left[-\log E_Y \ind\{\tilde\rho(X,Y)=0\}\right] \\
& \quad = E_X\left[-\log E_Y\ind\{\rho(X,Y)=\rho_Q(X)\}\right] .
\end{align*}
We moved the limit inside the expectations using first the monotone convergence theorem and then the dominated convergence theorem.
\end{IEEEproof}

\subsubsection{Proposition \ref{p:RPQD}}

Proposition \ref{p:RPQD} is an immediate consequence of the next two lemmas.  The proofs follow \cite{DemKon:Source:2002}[Thm.~2] with minor modifications.   Note that Proposition \ref{p:RPQD} and Lemma \ref{l:Lambda*} imply that $\Dmin= E \rho_Q(X)$ whenever the former is finite.  

\vskip 2ex
\begin{lem} \label{l:Lambda*1} If $W\in W(P,D)$, then $H(W\|\matsize{P}{Q}) \geq \Lambda^*(D)$.
\end{lem}
\vskip 2ex

\begin{IEEEproof}
Let $\psi:T\mapsto(-\infty,0]$ be measurable.  Then \cite{DemKon:Source:2002}
\[ H(\tilde Q\|Q) \geq E_{V \sim \tilde Q}\psi(V)-\log Ee^{\psi(Y)} \]
for any probability measure $\tilde Q$ on $T$.  Applying the previous inequality with $\psi(y):=\lambda\rho(x,y)$, for $\lambda\leq 0$, gives
\[ H(W(\cdot|x)\|Q) \geq \lambda E_{V \sim W(\cdot|x)} \rho(x,V) - \log Ee^{\lambda\rho(x,Y)} \]
where $W(\cdot|x)$ denotes the regular conditional distribution of $V$ given $U=x$ for $(U,V)\sim W$.
Taking expectations w.r.t.~$U$ and noting that $W\in W(P,D)$ gives
\[ H(W\|\matsize{P}{Q}) = E_{U\sim P} H(W(\cdot|U)\|Q) \geq \lambda D - \Lambda(\lambda) .\]
Optimizing over $\lambda\leq 0$ completes the proof.
\end{IEEEproof}

\vskip 2ex
\begin{lem} \label{l:Lambda*2} If $\Lambda^*(D) < \infty$, then there exists a $W\in W(P,D)$ with $H(W\|\matsize{P}{Q})=\Lambda^*(D)$.
\end{lem}
\vskip 2ex

\begin{IEEEproof}
The proof makes frequent use of Lemma \ref{l:Lambda*}.  
If $D \geq \Dave$, then $\Lambda^*(D) = 0$ and $W:= \matsize{P}{Q}$ achieves the equality.  If $\Dmin < D < \Dave$, then $W$ defined by
\[ \frac{dW}{d(\matsize{P}{Q})}(x,y) := \frac{e^{\lambda_D \rho(x,y)}}{Ee^{\lambda_D \rho(x,Y)}} \]
achieves the equality \cite{DemKon:Source:2002}, where $\lambda_D$ is uniquely chosen so that $\Lambda^*(D) = \lambda_D D-\Lambda(\lambda_D)$.  

Finally, if $D=\Dmin=E\rho_Q(X)$, then define $W$ by
\[ \frac{dW}{d(\matsize{P}{Q})}(x,y) := \frac{\ind\{y\in A(x)\}}{E\ind\{Y\in A(x)\}}  \]
where $A(x)=\{y:\rho(x,y)=\rho_Q(x)\}$.
Note that Lemma \ref{l:Lambda*} shows that $\Lambda^*(D) = E_{X}\left[-\log E_{Y}\ind\{Y\in A(X)\}\right]$ which we have assumed is finite, so the denominator is positive $P$-a.s.~and $W$ is well-defined.  It is easy to see that $W\in W(P,D)$ and that
\begin{align*} & H(W\|\matsize{P}{Q}) = E\left[\frac{dW}{d(\matsize{P}{Q})}(X,Y)\log\frac{dW}{d(\matsize{P}{Q})}(X,Y)\right] \\ & \quad = E\left[\frac{\ind\{Y\in A(X)\}}{E_Y\left[\ind\{Y\in A(X)\}\right]} \log \ind\{Y\in A(X)\}\right] \\ & \quad \quad - E\left[\frac{\ind\{Y\in A(X)\}}{E_Y\left[\ind\{Y\in A(X)\}\right]} \log E_{Y}\left[\ind\{Y\in A(X)\}\right]\right] \\
& \quad = 0 - E_{X}\left[\log E_{Y}\ind\{Y\in A(X)\}\right] = \Lambda^*(D) \end{align*}
which completes the proof.
\end{IEEEproof}

\subsection{Extensions to memory} \label{s:memory_proof}

Here we prove Proposition \ref{p:Rmix} and the claims in the text following Theorem \ref{t:mix}, including the existence of $R(\mathbb{P},\mathbb{Q},D)$, under the assumptions of Section \ref{s:memory}.  The stationarity and mixing properties of $\mathbb{Q}$ give $Q^n\ll Q_n \ll Q^n$, which proves \eqref{e:essinf}, and they give 
\begin{align}  & C^{-1}\int_{T^n}\int_{T^m} f(y_1^{n+m}) Q_m(dy_{n+1}^{n+m}) Q_n(dy_1^n) \notag \\ \quad & \leq \int_{T^{n+m}} f(y_1^{n+m}) Q_{n+m}(dy_1^{n+m}) \notag \\ & \quad \leq C\int_{T^n}\int_{T^m} f(y_1^{n+m}) Q_m(dy_{n+1}^{n+m}) Q_n(dy_1^n) \label{e:Qmix} \end{align}
for any function $f\geq 0$.  We make use of this property repeatedly.  Note that if $f$ factors, i.e., if $f(y_1^{n+m}) = g(y_1^n)h(y_{n+1}^{n+m})$ for $g,h\geq 0$, then \eqref{e:Qmix} becomes 
\begin{equation} \label{e:Qmixprod} C^{-1} E g(Y_1^n) E h(Y_1^m) \leq Ef(Y_1^{n+m}) \leq C E g(Y_1^n) E h(Y_1^m) . \end{equation}

This gives
\begin{align*} & C^{-1}\left[E_{Y_1^n}e^{n \lambda \rho_n(x_1^n,Y_1^n)}\right]\left[E_{Y_1^m}e^{m \lambda \rho_m(x_{n+1}^{n+m},Y_1^m)}\right] \\ & \quad \leq E_{Y_1^{n+m}}e^{(n+m) \lambda \rho_{n+m}(x_1^{n+m},Y_1^{n+m})} \\ & \quad \leq C\left[E_{Y_1^n}e^{n \lambda \rho_n(x_1^n,Y_1^n)}\right]\left[E_{Y_1^m}e^{m \lambda \rho_n(x_{n+1}^{n+m},Y_1^m)}\right] \end{align*}
which implies that
\begin{align} & \Lambda_n(\delta_{x_1^n},Q_n,n\lambda) + \Lambda_m(\delta_{x_{n+1}^{n+m}},Q_m,m\lambda) - \log C \notag \\ & \quad \leq \Lambda_{n+m}(\delta_{x_1^{n+m}},Q_{n+m},(n+m)\lambda) \notag \\ & \quad \leq \Lambda_n(\delta_{x_1^n},Q_n,n\lambda) + \Lambda_m(\delta_{x_{n+1}^{n+m}},Q_m,m\lambda) + \log C . \label{e:Lambdamix}  \end{align}
Replacing $x_k$ with $X_k$ and taking expected values gives
\begin{align}  & \Lambda_n(P_n,Q_n,n\lambda) + \Lambda_m(P_m,Q_m,m\lambda) - \log C \notag \\ & \quad \leq \Lambda_{n+m}(P_{n+m},Q_{n+m},(n+m)\lambda) \notag \\ & \quad \leq \Lambda_n(P_n,Q_n,n\lambda) + \Lambda_m(P_m,Q_m,m\lambda) + \log C .  \label{e:Lambdamix2} \end{align}

This final result implies several things.  First, it shows that if $\Lambda_n(P_n,Q_n,n\lambda)$ is finite (infinite) for some $n$, then it is finite (infinite) for all $n$.  It also shows that the sequence $\Lambda_n(P_n,Q_n,n\lambda)+\log C$ is subadditive, so the limit in the definition of $\Lambda_\infty$ exists.  In particular \cite{Kal:Foundations:2002}[Lemma 10.21],
\begin{align*} & \Lambda_\infty(\mathbb{P},\mathbb{Q},\lambda) := \lim_{n\to\infty} \frac{1}{n}\Lambda_n(P_n,Q_n,n\lambda) \\ & \quad = \inf_{n\geq N} \frac{1}{n}\left[\Lambda_n(P_n,Q_n,n\lambda) + \log C\right] \end{align*}
for any $N\geq 0$.  This gives
\begin{align*} & \Lambda_\infty^*(\mathbb{P},\mathbb{Q},D) \\ & \quad = \sup_{\lambda \leq 0}\left[\lambda D - \inf_{n\geq N} \frac{1}{n}\left[\Lambda_n(P_n,Q_n,n\lambda) + \log C\right]\right] \\
& \quad = \sup_{n\geq N}\left[\sup_{\lambda \leq 0}\left[\lambda D - \frac{1}{n}\Lambda_n(P_n,Q_n,n\lambda)\right] - \frac{\log C}{n}\right] \\ & \quad = \sup_{n\geq N} \left[\Lambda_n^*(P_n,Q_n,D) - \frac{\log C}{n}\right] . \end{align*}
The last equality follows from \eqref{e:LL*} which is easy to prove by moving the $1/n$ outside of the supremum and optimizing over $n \lambda$ instead of $\lambda$.  Since we always have
\begin{align*} & \Lambda_\infty^*(\mathbb{P},\mathbb{Q},D) = \sup_{\lambda\leq 0}\lim_{n\to\infty}\left[\lambda D - \frac{1}{n}\Lambda(P_n,Q_n,n \lambda)\right] \\ & \quad \leq \liminf_{n\to\infty} \sup_{\lambda\leq 0}\left[\lambda D - \frac{1}{n}\Lambda(P_n,Q_n,n \lambda)\right] \\ & \quad = \liminf_{n\to\infty} \Lambda_n^*(P_n,Q_n,D)  \end{align*}
we have also shown that 
\begin{align*} & \Lambda_\infty^*(\mathbb{P},\mathbb{Q},D)=\lim_{n\to\infty} \Lambda_n^*(P_n,Q_n,D) \\ & \quad = \lim_{n\to\infty} R_n(P_n,Q_n,D) := R(\mathbb{P},\mathbb{Q},D). \end{align*}  This completes the proof of Proposition \ref{p:Rmix} and shows that $R(\mathbb{P},\mathbb{Q},D)$ exists.

Lastly, \eqref{e:Lambdamix2} shows that
\[ \Lambda(P,Q,\lambda) - \log C \leq \frac{1}{n}\Lambda_n(P_n,Q_n,n \lambda) \leq \Lambda(P,Q,\lambda) + \log C  \]
so $\Lambda^*(P,Q,D) - \log C \leq \Lambda_n^*(P_n,Q_n,D) \leq \Lambda^*(P,Q,D) + \log C$.  This gives \eqref{e:DminDmin}.

\subsection{A large deviations result} \label{s:GE}

For appropriate values of $D$, the generalized AEP is essentially a large deviations result.  The next lemma summarizes what we need.  It is basically a corollary of the G\"artner-Ellis Theorem.  Note that $\Lambda$ and $\Lambda^*$ are redefined in this section.

\vskip 2ex
\begin{lem} \label{l:GE} Let $(Z_n)_{n\geq 1}$ be a sequence of nonnegative, real-valued random variables such that \[ \Lambda(\lambda) := \lim_{n\to\infty} \frac{1}{n}\log Ee^{n \lambda Z_n} \ \text{ exists} \]
for all $\lambda\in\mathbb{R}$.  Define $\Lambda^*(D):=\sup_{\lambda\leq 0}\left[\lambda D - \Lambda(\lambda)\right]$.  Then
\[ \limsup_{n\to\infty} \frac{1}{n}\log \Prob\{Z_n\leq D\} \leq -\Lambda^*(D) \]
for all $D$.
Furthermore, if $\Lambda^*$ is strictly convex on $(a,b)$, then
\[ \lim_{n\to\infty} \frac{1}{n}\log \Prob\{Z_n\leq D\} = -\Lambda^*(D) \]
for all $D\in (a,b]$.
\end{lem}
\vskip 2ex

\begin{IEEEproof}
For any $\lambda\leq 0$, $\Prob\{Z_n\leq D\}\leq Ee^{n \lambda (Z_n-D)}$, so
\begin{align*} & \limsup_{n\to\infty} \frac{1}{n}\log \Prob\{Z_n \leq D\} \\ & \quad \leq -\lambda D + \limsup_{n\to\infty} \frac{1}{n}\log E^{n \lambda Z_n}  = -[\lambda D - \Lambda(\lambda)] . \end{align*}
Optimizing over $\lambda\leq 0$ gives the upper bound.  
  
Suppose $\Lambda^*$ is strictly convex on $(a,b)$.  Since $\Lambda^*$ is nonnegative and decreasing, $\Lambda^*$ must be finite and positive on $(a,b)$.  The finiteness implies that $\Lambda$ is finite on $(-\infty,0]$.  We will first show that
\begin{equation} \label{e:conj} \Lambda^*(D) = \sup_{\lambda\in\mathbb{R}}\left[\lambda D - \Lambda(\lambda)\right] \quad \quad D\leq b . \end{equation}
It is easy to see that $\Lambda$ is increasing and convex with $\Lambda(0) = 0$, so we can choose a $0 \leq D' \leq \infty$ with $\Lambda(\lambda)\geq \lambda D'$ for all $\lambda\in\mathbb{R}$.  If $D'=\infty$, then $\Lambda(\lambda)=\infty$ for $\lambda >0$ and \eqref{e:conj} holds for all $D$.  If $D'$ is finite and $D \leq D'$, then $\lambda D - \Lambda(\lambda) \leq \lambda D' - \Lambda(\lambda) \leq 0$ for all $\lambda > 0$, so \eqref{e:conj} holds for all $D \leq D'$.  The same inequality gives $\Lambda^*(D')=0$, so $b \leq D'$.

Now we will prove the lower bound.  If $\Lambda$ is finite in some neighborhood of zero, then the lemma follows immediately from the G\"artner-Ellis Theorem as stated in \cite{Hol:Large:2000}[Thm.~V.6].  If this is not the case, then we need to slightly modify the sequence $(Z_n)$ before applying the theorem.

Fix $D\in (a,b]$ and choose $0 < \epsilon < D-a$.
Let $(\hat Z_n)_{n\geq 1}$ be a sequence of nonnegative, real-valued r.v.'s with distribution $\hat P_n(\cdot) := \Prob\{\hat Z_n \in \cdot\}$ defined by
\[ \frac{d\hat P_n}{d P_n}(z) := \frac{e^{-n \epsilon z}}{Ee^{-n \epsilon Z_n}}  \quad \quad z \geq 0  \]
where $P_n(\cdot):=\Prob\{Z_n\in\cdot\}$.  We have
\begin{align*} & \log \Prob\{Z_n\leq D\} \geq \log P_n((D-\epsilon,D)) \\ & \quad = \log \int_{D-\epsilon}^D \frac{Ee^{-n \epsilon Z_n}}{e^{-n \epsilon z}} \hat P_n(dz) \\
& \quad \geq \log Ee^{-n \epsilon Z_n} + n \epsilon (D-\epsilon) + \log \hat P_n((D-\epsilon,D)) . \end{align*}
Taking limits gives
\begin{align} & \liminf_{n\to\infty} \frac{1}{n} \log \Prob\{Z_n\leq D\} \notag \\ & \quad \geq \Lambda(-\epsilon) + \epsilon D - \epsilon^2 + \liminf_{n\to\infty} \frac{1}{n} \log \hat P_n((D-\epsilon,D)) .  \label{e:p2phat} \end{align}

We want to apply the G\"artner-Ellis Theorem to the sequence $(\hat P_n)_{n\geq 1}$.  Note that 
\[ Ee^{n \lambda \hat Z_n} = \int e^{n \lambda z} \frac{e^{-n \epsilon z}}{Ee^{-n \epsilon Z_n}} P_n(dz) = \frac{Ee^{n (\lambda-\epsilon) Z_n}}{Ee^{-n \epsilon Z_n}} \]
so
\[ \hat\Lambda(\lambda) := \lim_{n\to\infty} \frac{1}{n}\log Ee^{n \lambda \hat Z_n} = \Lambda(\lambda-\epsilon) - \Lambda(-\epsilon) \]
exists and is finite for all $\lambda \leq \epsilon$.  In particular, it is finite in a neighborhood of $0$.  Note also that 
\begin{align*} & \hat\Lambda^*(x) := \sup_{\lambda\in\mathbb{R}}\left[\lambda x - \hat\Lambda(\lambda)\right] = \sup_{\lambda\in\mathbb{R}} \left[ (\lambda+\epsilon) x - \hat\Lambda(\lambda+\epsilon)\right] \\
& = \sup_{\lambda\in\mathbb{R}} \left[\lambda x - \Lambda(\lambda)\right] + \epsilon x + \Lambda(-\epsilon) = \Lambda^*(x) + \epsilon x + \Lambda(-\epsilon) \end{align*}
for any $x \leq b$.  So $\hat\Lambda^*$ is also strictly convex on $(a,b)$ and the slope of any supporting line to $\hat\Lambda^*$ at a point in $(a,b)$ is strictly less than $\epsilon$.  In particular, the slope of such a point is in the interior of the domain where $\hat\Lambda$ is finite.  So the assumptions of the G\"artner-Ellis Theorem are satisfied and
\begin{align*} & \liminf_{n\to\infty} \frac{1}{n}\log \hat P_n((D-\epsilon,D)) \geq -\inf_{x\in(D-\epsilon,D)} \hat \Lambda^*(x) \\ & \quad = -\inf_{x\in(D-\epsilon,D)} \left[\Lambda^*(x) + \epsilon x + \Lambda(-\epsilon)\right] \\ & \quad \geq -\inf_{x\in(D-\epsilon,D)} \left[\Lambda^*(x) + \epsilon D + \Lambda(-\epsilon)\right] \\ & \quad = -\Lambda^*(D) - \epsilon D - \Lambda(-\epsilon) . 
\end{align*}
Combining this with \eqref{e:p2phat} gives
\[ \liminf_{n\to\infty}  \frac{1}{n} \log \Prob\{Z_n\leq D\} \geq -\Lambda^*(D) - \epsilon^2 . \]
Since $\epsilon$ was arbitrary, this completes the proof.
\end{IEEEproof}

\vskip 2ex
\begin{lem} \label{l:LandR}
Let $Z$ be a real-valued, nonnegative random variable.  Define $\Lambda^*(D) := \sup_{\lambda \leq 0} [\lambda D - \log Ee^{\lambda Z}]$.  Then
\[ \log \Prob\{Z \leq D\} \leq -\Lambda^*(D) \]
with equality for $D\leq \essinf Z$.
Furthermore, $\log\Prob\{Z \leq D\}$ is finite if and only if $-\Lambda^*(D)$ is finite.
\end{lem}
\vskip 2ex

\begin{IEEEproof}
For any $\lambda \leq 0$, $\log \Prob\{Z \leq D\} \leq -[\lambda D - \log Ee^{\lambda Z}]$.  Optimizing over $\lambda\leq 0$ gives the first bound.  Suppose $D\leq\essinf Z$ so that $Z-D\relas{\geq} 0$. In this case
\begin{align*} & \Prob\{Z\leq D\} = \Prob\{Z=D\} = \lim_{\lambda\to -\infty} Ee^{\lambda(Z-D)} \\ & \quad = \inf_{\lambda\leq 0} Ee^{\lambda(Z-D)} \end{align*}
and
\begin{align*} & \log\Prob\{Z\leq D\} = \inf_{\lambda\leq 0} \left[\log Ee^{\lambda Z} - \lambda D\right] 
= -\Lambda^*(D) . \end{align*}
Of course, if $D > \essinf Z$, then $-\infty < \log\Prob\{Z\leq D\} \leq -\Lambda^*(D) \leq 0$, and everything is finite.
\end{IEEEproof}

\vskip 2ex
\begin{cor} \label{c:R}
Lemma \ref{l:GE} holds if $n^{-1}\log \Prob\{Z_n \leq D\}$ is replaced by $-\Lambda_n^*(D)$, where
\[ \Lambda_n^*(D) := \frac{1}{n}\sup_{\lambda \leq 0}\left[\lambda D - \log E e^{\lambda Z_n}\right] . \]
\end{cor}
\vskip 2ex

\begin{IEEEproof}
$-\Lambda_n^*(D) \leq -[n \lambda D - \log E^{n \lambda Z_n}]/n$.
Taking limits and optimizing over $\lambda\leq 0$ gives the upper bound
\[ \limsup_{n\to\infty} -\Lambda_n^*(D) \leq -\Lambda^*(D) .\]

Lemma \ref{l:LandR} shows that 
\[ \liminf_{n\to\infty} -\Lambda_n^*(D) \geq \liminf_{n\to\infty} \frac{1}{n}\log\Prob\{Z_n\leq D\} , \]
which gives the lower bound in the second part of Lemma \ref{l:GE}.
\end{IEEEproof}

\subsection{The generalized AEP}

Now we will prove the main theorems in the text.  We focus on the more general setting with memory described in Section \ref{s:memory} since this includes the memoryless situation as a special case.  The main idea is to fix a typical realization $(x_n)_{n\geq 1}$ of $(X_n)_{n\geq 1}$ and then analyze the behavior of the sequence of r.v.'s $(Z_n)_{n\geq 1}$, where 
\begin{equation} \label{e:Zn} Z_n := \rho_n(x_1^n,Y_1^n) := \frac{1}{n}\sum_{k=1}^n \rho(x_k,Y_k) \end{equation}
and where $(Y_n)_{n\geq 1}$ has distribution $\mathbb{Q}$.  Using this terminology, 
\[ L_n(x_1^n,Q_n,D) = -\frac{1}{n}\log\Prob\{Z_n\leq D\} \]
and
\begin{align*} & R_n(\delta_{x_1^n},Q_n,D) = \Lambda^*_n(\delta_{x_1^n},Q_n,D) \\ & \quad := \frac{1}{n}\sup_{\lambda\leq 0}\left[\lambda D - \log Ee^{\lambda Z_n}\right]  . \end{align*}

The proof proceeds in several stages.   Proposition \ref{p:Rmix} allows us to use $\Lambda^*_\infty(\mathbb{P},\mathbb{Q},D)$ instead of $R_\infty(\mathbb{P},\mathbb{Q},D)$.  We first prove the lower bound
\begin{equation} \liminf_{n\to\infty} L_n(X_1^n,Q_n,D) \relas{\geq} \Lambda^*_\infty(\mathbb{P},\mathbb{Q},D)  \label{e:LB} \end{equation}
for all $D$.   Then we prove the upper bound
\begin{equation} \limsup_{n\to\infty} L_n(X_1^n,Q_n,D) \relas{\leq}  \Lambda^*_\infty(\mathbb{P},\mathbb{Q},D)  \label{e:UB} \end{equation}
separately for the cases $D < \Dmin(P,Q)$, $D > \Dave(P,Q)$ and $\Dmin(P,Q) < D \leq \Dave(P,Q)$.  The case $D=\Dmin(P,Q)$ can be pathological in certain situations.  For these situations we characterize the pathology as described in Theorem \ref{t:iff} (extended to the situation with memory).   
Note that even in the pathological situation when the limit does not exist, there is a subsequence along which the upper bound in \eqref{e:UB} holds.  This gives Theorem \ref{t:genAEP} (extended to the situation with memory).
Finally, Lemma \ref{l:LandR} allows us to replace $L_n(X_1^n,Q_n,D)$ with $R_n(\delta_{x_1^n},Q_n,D)$ along the lines of Corollary \ref{c:R}, even in the pathological situation.

\subsubsection{The lower bound} \label{s:LB}

\eqref{e:Lambdamix} shows that we can apply the subadditive ergodic theorem \cite{Kal:Foundations:2002}[Theorem 10.22] to \[ \Lambda_n(\delta_{X_1^n},Q_n,n\lambda) + \log C \] for $\lambda \leq 0$ or to \[ -\Lambda_n(\delta_{X_1^n},Q_n,n\lambda)+\log C \] for $\lambda \geq 0$ (so that everything is bounded above by $\log C$) to get
\begin{equation} \lim_{n\to\infty} \frac{1}{n}\Lambda_n(\delta_{X_1^n},Q_n,n\lambda) \relas{=} \Lambda_\infty(\mathbb{P},\mathbb{Q},D) . \label{e:Lambdalim} \end{equation}
The right side is a constant because the limit is shift-invariant and the source is ergodic.
Since $\Lambda_n$ is increasing in $\lambda$, we can choose the exceptional set independently of $\lambda$.  

Choosing $(x_n)_{n\geq 1}$ so that \eqref{e:Lambdalim} holds and defining $(Z_n)_{n\geq 1}$ as in \eqref{e:Zn} allows us to apply the first part of Lemma \ref{l:GE} to get the lower bound \eqref{e:LB}.  Note that Corollary \ref{c:R} gives the same lower bound for $R_n(\delta_{X_1^n},Q_n,D)$.

\subsubsection{The upper bound when $D < \Dmin$ or $D > \Dave$}

When $\Lambda^*(\mathbb{P},\mathbb{Q},D)=\infty$, the lower bound \eqref{e:LB} implies the upper bound \eqref{e:UB}.  Note that this includes all $D < \Dmin(\mathbb{P},\mathbb{Q})$ and possibly some situations where $D=\Dmin(\mathbb{P},\mathbb{Q})$.

If $\Dave(P,Q)$ is finite and $D > \Dave(P,Q)$, then Chebyshev's inequality and the ergodic theorem give 
\begin{align*} & L_n(X_1^n,Q_n,D) = -\frac{1}{n}\log\left[1-Q_n\left\{y_1^n:\rho_n(X_1^n,y_1^n) > D\right\}\right] \\ & \quad \leq -\frac{1}{n}\log\left[1-\frac{1}{D}E_{Y_1^n}\rho_n(X_1^n,Y_1^n)\right] \\ & \quad \relas{\to} 0 \leq \Lambda^*_\infty(\mathbb{P},\mathbb{Q},D) \end{align*}
as $n\to\infty$, since $E_{Y_1^n}\rho_n(X_1^n,Y_1^n)\relas{\to} \Dave(P,Q) < D$.  This gives the upper bound \eqref{e:UB} for the case $D > \Dave(P,Q)$.

\subsubsection{The upper bound when $\Dmin < D \leq \Dave$}

Assume that $\Dmin:=\Dmin(\mathbb{P},\mathbb{Q}) < D \leq \Dave(P,Q):=\Dave$.  If $\Lambda^*_\infty(\mathbb{P},\mathbb{Q},\cdot)$ is known to be strictly convex on $(\Dmin,\Dave)$, then we could apply the second part of Lemma \ref{l:GE} in the same manner as Section \ref{s:LB} to get the upper bound on $(\Dmin,\Dave]$.  Unfortunately, we were unable to find a simple proof of this strict convexity.  Instead we will apply Lemma \ref{l:GE} to an approximating sequence of random variables $(\hat Z_n)_{n\geq 1}$.

Fix $m\in\mathbb{N}$.  Let $\hat{\mathbb{Q}}$ denote the distribution of a random process $(\hat Y_n)_{n\geq 1}$ taking values in $T$ with the property that $\hat Y_{(n-1)m+1}^{nm}$ has distribution $Q_m$ and is independent of all the other $\hat Y_k$'s.  We use $\hat Q_n$ to denote the distribution of $\hat Y_1^n$.  
If $n=m\ell+r$, $1 \leq r \leq m$, then $\hat Q_n = \left(\times_{k=1}^\ell Q_m\right)\times Q_r$ and
\begin{equation} C^{-\ell}\hat Q_n(A) \leq Q_n(A) \leq C^{\ell}\hat Q_n(A) \label{e:hatQmix} . \end{equation} 
The next Lemma summarizes how $\hat{\mathbb{Q}}$ behaves in our context.

\vskip 2ex
\begin{lem} \label{l:hatLambda}  Fix $m\in\mathbb{N}$ and define $\hat{\mathbb{Q}}$ as above.  Then 
\begin{align} & \Lambda_\infty(\delta_{X_1^\infty},\hat{\mathbb{Q}},\lambda) := \lim_{n\to\infty} \frac{1}{n}\Lambda_n(\delta_{X_1^n},\hat Q_n,n\lambda) \notag \\ & \quad  = \frac{1}{m}\Lambda_m(P_m(\cdot|\mathcal{I}),Q_m,m\lambda) \label{e:hatLambda} \end{align}
exists and has the above representation for all $\lambda\in\mathbb{R}$ with probability 1, where $P_m(\cdot|\mathcal{I})$ is a random probability distribution on $S^m$ depending only on the sequence $X_1^\infty$.   Furthermore, 
\[ \Lambda^*_\infty(\delta_{X_1^\infty},\hat{\mathbb{Q}},D) := \sup_{\lambda\leq 0}\left[\lambda D - \Lambda_\infty(\delta_{X_1^\infty},\hat{\mathbb{Q}},\lambda)\right]  \]
is strictly convex in $D$ on $(\Dmin,\Dave)$ and 
\begin{align} & \Lambda^*_\infty(\delta_{X_1^\infty},\hat{\mathbb{Q}},D) - \frac{\log C}{m} \leq \Lambda^*_\infty(\mathbb{P},\mathbb{Q},D) \notag \\ & \quad \leq \Lambda^*_\infty(\delta_{X_1^\infty},\hat{\mathbb{Q}},D) + \frac{\log C}{m}  \label{e:hatLambda*} \end{align}
for all $D$ with probability 1.
\end{lem}
\vskip 2ex

\begin{IEEEproof}
To simplify notation, fix $\lambda$ and define the r.v. \[ \hat\Lambda_n:=\Lambda(\delta_{X_1^n},\hat Q_n,n\lambda) . \]  We will first show that the convergence of $\hat\Lambda_n/n$ is a.s.~determined by the convergence of the subsequence $\hat\Lambda_{m\ell}/(m\ell)$ as $\ell\to\infty$.

The ergodic theorem gives
\begin{equation} \frac{1}{n}\sum_{k=1}^n \Lambda(\delta_{X_k},Q,\lambda) \relas{\to} \Lambda(P,Q,\lambda) . \label{e:proofQhat1} \end{equation}
Analogous to the arguments in Section \ref{s:memory_proof},
\begin{equation} \frac{1}{n}\hat\Lambda_n \in \frac{1}{n}\sum_{k=1}^n \Lambda(\delta_{X_k},Q,\lambda) \pm \log C  \label{e:proofQhat2} . \end{equation}
If $\Lambda(P,Q,\lambda)$ is infinite, then \eqref{e:proofQhat1} and \eqref{e:proofQhat2} show that $\lim_n \hat\Lambda_n/n$ exists and is infinite a.s.  In particular, $\lim_n \hat\Lambda_n/n \relas{=} \lim_{\ell} \hat\Lambda_{m\ell}/(m\ell)$.  

If $\Lambda(P,Q,\lambda)$ is finite, then \eqref{e:proofQhat1} shows that
\[ \frac{1}{n}\Lambda(\delta_{X_n},Q,\lambda) \relas{\to} 0 \]
which implies that
\begin{equation} \label{e:proofQhat3} \frac{1}{n}\Lambda_r(\delta_{X_{n-r+1}^n},Q_r,r\lambda) \relas{\to} 0 \end{equation}
for each $r$; see \eqref{e:Lambdamix}.   Writing $n=m\ell+r$ for $1\leq r \leq m$, the block-independence property of $\hat{\mathbb{Q}}$ gives
\[ \hat\Lambda_n = \hat\Lambda_{m\ell} + \Lambda_r(\delta_{X_{\ell m + 1}^n},Q_r,r\lambda) . \]
Combining this with \eqref{e:proofQhat3} shows that $\hat\Lambda_n/n$ has a.s.~the same asymptotic behavior as $\hat\Lambda_{m\ell}/(m\ell)$.

We will now analyze the limiting behavior of $\hat\Lambda_{m\ell}/(m\ell)$.  The block-independence property of $\hat{\mathbb{Q}}$ gives
\begin{equation} \frac{1}{m\ell}\hat\Lambda_{m\ell} = \frac{1}{m\ell} \sum_{k=1}^{\ell} \Lambda_m(\delta_{X_{m(k-1)+1}^{mk}},Q_m,m\lambda) .  \label{e:proofQhat4} \end{equation}
The sequence $(X_{m(\ell-1)+1}^{m\ell})_{\ell\geq 1}$ 
of disjoint $m$-blocks from $(X_n)_{n\geq 1}$ is stationary 
(but not necessarily ergodic), so the ergodic theorem 
\cite[Theorem 10.6]{Kal:Foundations:2002}
gives
\begin{align} & \lim_{\ell\to\infty} \frac{1}{\ell}\sum_{k=1}^{\ell} \Lambda_m(\delta_{X_{(k-1)m+1}^{km}},Q_m,m\lambda) \notag \\ & \quad \relas{=} E\left[\Lambda_m(\delta_{X_1^m},Q_m,m\lambda)\bigl|\mathcal{I}\right] \label{e:hatLambda:1}  \end{align}
where $\mathcal{I}$ is the shift invariant $\sigma$-field for the sequence $(X_{m(\ell-1)m+1}^{m\ell})_{\ell\geq 1}$.  Letting $P_m(\cdot|\mathcal{I})$ denote the regular conditional distribution of $X_1^m$ given $\mathcal{I}$, the right side of \eqref{e:hatLambda:1} is $\Lambda_m(P_m(\cdot|\mathcal{I}),Q_m,m\lambda)$.  

Combining \eqref{e:proofQhat4} and \eqref{e:hatLambda:1} and recalling our discussion about the subsequence $(m\ell)_{\ell\geq 1}$ shows that \eqref{e:hatLambda} holds a.s.~for each specific $\lambda$.
Since $\Lambda_n$ is increasing and since $\mathcal{I}$ does not depend on $\lambda$, we can choose the exceptional set independently of $\lambda$.  This implies that the corresponding $\Lambda_\infty^*$ is a.s.~well-defined and the exceptional set does not depend on $D$. 

Two applications of the ergodic theorem show that
\begin{align} & \Dave \relas{=} \lim_{n\to\infty} \frac{1}{n}\sum_{k=1}^n E_{Y_1}\rho(X_k,Y_1) \notag \\ & \quad = \lim_{\ell\to\infty} \frac{1}{m\ell}\sum_{k=1}^\ell \sum_{j=1}^m E_{Y_1}\rho(X_k,Y_1) 
\notag \\ & \quad = \frac{1}{\ell}\sum_{k=1}^\ell E_{Y_1^m}\rho_m(X_{(k-1)m+1}^{km},Y_1^m) \notag \\ & \quad \relas{=} E\left[E_{Y_1^m}\rho_m(X_1^m,Y_1^m)\bigl|\mathcal{I}\right] \notag \\ & \quad = E_{X_1^m\sim P_m(\cdot|\mathcal{I})}\left[E_{Y_1^m}\rho_m(X_1^m,Y_1^m)\right] . \label{e:QhatDave} \end{align}
An identical argument, combined with \eqref{e:essinf}, gives
\begin{equation} \Dmin \relas{=}  E_{X_1^m\sim P_m(\cdot|\mathcal{I})}\left[\essinf_{Y_1^m}\rho_m(X_1^m,Y_1^m)\right] . \label{e:QhatDmin} \end{equation}

Because of the representation on the right side of \eqref{e:hatLambda}, we can apply Lemma \ref{l:Lambda*} with $S=S^m$, $T=T^m$, $\rho=\rho_m$, $X\sim P_m(\cdot|\mathcal{I})$, and $Y\sim Q_m$ to see that $\Lambda_\infty^*(\delta_{X_1^\infty},\hat{\mathbb{Q}},\cdot)$ is strictly convex on $(\Dmin,\Dave)$ a.s.  Identifying the $\Dmin$ and $\Dave$ from Lemma \ref{l:Lambda*} with $\Dmin$ and $\Dave$ here follows from \eqref{e:QhatDmin} and \eqref{e:QhatDave} above.

Finally, analogous to the arguments in Section \ref{s:memory_proof}, \eqref{e:hatQmix} gives
\begin{align*} & \Lambda_n(\delta_{x_1^n},\hat Q_n,n\lambda) - \frac{n}{m}\log C \leq \Lambda_n(\delta_{x_1^n},Q_n,n\lambda) \\ & \quad \leq \Lambda_n(\delta_{x_1^n},\hat Q_n,n\lambda) + \frac{n}{m}\log C . \end{align*} 
Combining this with \eqref{e:hatLambda}  and \eqref{e:Lambdalim} gives \eqref{e:hatLambda*}. 
\end{IEEEproof}

Returning to the main argument, fix a realization $(x_n)_{n\geq 1}$ of $(X_n)_{n\geq 1}$ so that everything holds in Lemma \ref{l:hatLambda}.  Define the sequence of random variables $(Z_n)_{n\geq 1}$ and $(\hat Z_n)_{n\geq 1}$ by $Z_n := \rho_n(x_1^n,Y_1^n)$ and $\hat Z_n := \rho_n(x_1^n,\hat Y_1^n)$.  \eqref{e:hatQmix} shows that
\begin{align*} &  L_n(x_1^n,Q_n,D) = -\frac{1}{n}\log Q_n(B_n(x_1^n,D)) \\ & \quad \leq -\frac{1}{n}\log \hat Q_n(B_n(x_1^n,D)) + \frac{\log C}{m} \\ & \quad = -\frac{1}{n}\log\Prob\{\hat Z_n \leq D\} + \frac{\log C}{m} .\end{align*}
Lemma \ref{l:hatLambda} lets us apply the second part of Lemma \ref{l:GE} to the right side to get
\begin{align*} & \limsup_{n\to\infty} L_n(x_1^n,Q_n,D) \leq \Lambda_\infty^*(\delta_{X_1^\infty},\hat{\mathbb{Q}},D) + \frac{\log C}{m} \\ & \quad \leq \Lambda_\infty^*(\mathbb{P},\mathbb{Q},D) + 2\frac{\log C}{m} \end{align*}
for all $D\in(\Dmin,\Dave]$.
The final inequality comes from \eqref{e:hatLambda*}.  Since $m$ was arbitrary and since $(x_n)_{n\geq 1}$ was a.s.~arbitrary, we have established the upper bound \eqref{e:UB} for the case $\Dmin < D \leq \Dave$.

\subsubsection{The case $D=\Dmin$}

We have established the lower bound \eqref{e:LB} for all $D$ and the upper bound \eqref{e:UB} for all $D$ except for the case when $D=\Dmin:=\Dmin(P,Q)$ and $\Lambda_\infty^*(\mathbb{P},\mathbb{Q},\Dmin) < \infty$.  We analyze that situation here.  To simplify notation, we will suppress the dependence on $\mathbb{P}$ and $\mathbb{Q}$ whenever it is clear from the context.

Define
\[ A_n(x_1^n) := \left\{y_1^n\in T^n : \rho_n(x_1^n,y_1^n) = \essinf_{Y_1^n} \rho_n(x_1^n,Y_1^n)\right\} . \] 
Because of \eqref{e:essinf}, 
\[ Q_{n+m}\left(A_{n+m}(x_1^{n+m})\right) = Q_{n+m}\left(A_n(x_1^n)\times A_m(x_{n+1}^{n+m})\right) \]
and the mixing properties of $\mathbb{Q}$ give
\begin{align*} & -\log Q_{n+m}\left(A_{n+m}(x_1^{n+m})\right) + \log C \\ & \quad \leq \left[-\log Q_n\left(A_n(x_1^n)\right) + \log C\right] \\ & \quad \quad + \left[- \log Q_m\left(A_m(x_{n+1}^{n+m})\right) + \log C\right] . \end{align*}
Lemma \ref{l:Lambda*} shows that
\[ E\left[-\log Q_n(A_n(X_1^n))\right] = n\Lambda_n^*(P_n,Q_n,\Dmin)  \]
which we assume is finite, so we can apply the subadditive ergodic theorem and Proposition \ref{p:Rmix} to get
\begin{equation} \label{e:QA} \lim_{n\to\infty} -\frac{1}{n} \log Q_n(A_n(X_1^n)) \relas{=} \Lambda_\infty^*(\mathbb{P},\mathbb{Q},\Dmin) . \end{equation}
   
Note that if $\rho_Q(X_1)$ is a.s.~constant, then $Q_n(A_n(X_1^n)) \relas{=} Q_n(B_n(X_1^n,\Dmin))$ and \eqref{e:QA} gives the upper bound.

Now suppose $\rho_Q(X_1)$ is not a.s.~constant (and $D=\Dmin$ and $\Lambda^*(\Dmin)<\infty$).  This is the only pathological situation where the upper bound does not hold.  Our analysis makes use of recurrence properties for random walks with stationary and ergodic increments.\footnote{$(W_n)_{n\geq 0}$ is a random walk with stationary and ergodic increments \cite{Ber:Random:1979} if $W_0:= 0$ and $W_n := \sum_{k=1}^n U_k$, $n\geq 1$, for some stationary and ergodic sequence $(U_n)_{n\geq 1}$.}  What we need is summarized in the following lemma:

\vskip 2ex
\begin{lem} \label{l:walk} 
Let $(U_n)_{n\geq 1}$ be a real-valued stationary and ergodic process and define $W_n:=\sum_{k=1}^n U_k$, $n\geq 1$.  If $EU_1 = 0$ and $\Prob\{U_1 \neq 0\} > 0$, then
$\Prob\left\{W_n > 0 \text{ i.o.}\right\} > 0$ and  $\Prob\left\{W_n \geq 0 \text{ i.o.}\right\} = 1$.
\end{lem} 
\vskip 2ex

\begin{IEEEproof}
Define $W_0 := 0$.  $(W_n)_{n\geq 0}$ is a random walk with stationary and ergodic increments. \cite{Kes:Sums:1975} shows that $\{ \liminf_n n^{-1} W_n > 0\}$ and $\{W_n\to\infty\}$ differ by a null set.  The ergodic theorem gives $\Prob\{ n^{-1}W_n \to 0\}=1$, so $\Prob\{W_n\to\infty\}=0$.  Similarly, by considering the process $-W_n$, we see that $\Prob\{W_n\to -\infty\}=0$.

Now $\{|W_n|\to\infty\}$ is invariant and must have probability $0$ or $1$.  If it has probability $1$, then since we cannot have $W_n\to\infty$ or $W_n\to-\infty$ we must have $W_n$ oscillating between increasingly larger positive and negative values, which means $\Prob\{W_n > 0 \text{ i.o.}\}=1$ and completes the proof.
  
Suppose $\Prob\{|W_n|\to\infty\}=0$.  Define
\[ N(A) := \sum_{n\geq 0} \ind\{W_n\in A\} , \quad A \subset \mathbb{R}, \]
to be the number of times the random walk visits the set $A$.   \cite{Ber:Random:1979}[Corollary 2.3.4] shows that either $N(J) < \infty$ a.s.~for all bounded intervals $J$ or $\{N(J)=0\}\cup\{N(J)=\infty\}$ has probability 1 for all intervals $J$ (open or closed, bounded or unbounded, but not a single point).  By assumption 
$|W_n|\not\to\infty$, so we can rule out the first possibility.  Since $\Prob\{W_0=0\}=1$, we see that for any interval $J$ containing $\{0\}$ we must have $\Prob\{N(J)=\infty\}=1$.  In particular, taking $J:=[0,\infty)$ shows that $\Prob\{W_n \geq 0 \text{ i.o.}\}=1$.  Similarly, taking $J:=(0,\infty)$ shows that $\Prob\{W_n > 0 \text{ i.o.}\} = \Prob\{N(J)=\infty\} = \Prob\{N(J) > 0\} \geq \Prob\{U_1 > 0\} > 0$.
\end{IEEEproof}

Returning to the main argument, 
\begin{align} & L_n(X_1^n,Q_n,\Dmin) \notag \\ & \quad \geq -\frac{1}{n}\log Q_n\left\{y_1^n : \frac{1}{n}\sum_{k=1}^n \rho_Q(X_k) \leq \Dmin\right\} 
\notag \\ & \quad = \begin{cases} 0 & \text{if $\sum_{k=1}^n \rho_Q(X_k) \leq n\Dmin$} \\ \infty & \text{if $\sum_{k=1}^n \rho_Q(X_k) > n\Dmin$} \end{cases} \notag \\ & \quad 
= \begin{cases} 0 & \text{if $W_n \leq 0$} \\ \infty & \text{if $W_n > 0$} \label{e:genAEP_1} \end{cases} , \end{align}
where $W_n := \sum_{k=1}^n (\rho_Q(X_k)-\Dmin)$.  Lemma \ref{l:walk} shows that $\Prob\{W_n > 0 \text{ i.o.}\} > 0$.  This and \eqref{e:genAEP_1} prove \eqref{e:genAEP_Dmin_infty}.  

Lemma \ref{l:walk} also shows that $\Prob\{W_n \leq 0 \text{ i.o.}\}=1$.  Let $(N_m)_{m\geq 1}$ be the (a.s.)~infinite, random subsequence of $(n)_{n\geq 1}$ such that $W_n \leq 0$.  Note that
\[ \frac{1}{N_m}\sum_{k=1}^{N_m} \rho_Q(X_k) \leq \Dmin \]
so
\begin{align} & L_{N_m}(X_1^{N_m},Q_{N_m},\Dmin) \notag \\ & \quad \leq -\frac{1}{N_m}\log Q_{N_m}{\textstyle \Bigl(B_{N_m}\bigl(X_1^{N_m},\frac{1}{N_m}\sum_{k=1}^{N_m}\rho_Q(X_k)\bigr)\Bigr) }
\notag \\ & \quad = -\frac{1}{N_m}\log Q_{N_m}(A_{N_m}(X_1^{N_m})) . 
\label{e:genAEP_2} \end{align}
Now, the final expression in \eqref{e:genAEP_2} is a.s.~finite because $E[-\log Q_n(A_n(X_1^n))] = n\Lambda_n^*(\Dmin) < \infty$.  This proves \eqref{e:genAEP_Dmin_finite} and shows that $(N_m)_{m\geq 1}$ satisfies the claims of the theorem, including \eqref{e:genAEP_essinf}.  Letting $m\to\infty$ in \eqref{e:genAEP_2} and using \eqref{e:QA} gives \eqref{e:genAEP_Dmin_lim}, the upper bound along the sequence $(N_m)_{m\geq 1}$.  Note that it also shows that the $\liminf_n$ is a.s.~$\Lambda^*_\infty$ even in this pathological case.

\subsubsection{Replacing $L_n$ with $R_n$}

Defining $Z_n := \rho_n(x_1^n,Y_1^n)$, Proposition \ref{p:RPQD} and Lemma \ref{l:LandR} show that
\[ R_n(\delta_{x_1^n},Q_n,D) = \Lambda^*_n(\delta_{x_1^n},Q_n,D) \leq  L_n(x_1^\infty,Q_n,D)   \]
and that $R_n$ and $L_n$ are finite (infinite) together.  Since we have already established that
$L_n(X_1^\infty,Q_n,D)$ and $R_n(\delta_{X_1^n},Q_n,D)$ have the same lower bound \eqref{e:LB}, we can use the above bound to squeeze $R_n$ when ever $\lim_n L_n$ exists.  

In the only pathological situation where the limit does not exist, $L_n$ converges along the subsequence where it is finite, so $R_n$ converges along that subsequence also.  But as we noted above, $L_n$ and $R_n$ have the same subsequence where they are finite.  

%
%

\section*{Acknowledgments}
I want to thank I.~Kontoyiannis, M.~Madiman and an anonymous reviewer for many useful comments and corrections, and I.~Kontoyiannis for invaluable advice and for suggesting the problems that led to this paper.

\bibliographystyle{IEEEtranS}
\bibliography{IEEEabrv,mathbib}

\end{document}

%% file: mathheader.tex
\usepackage{amsmath}
\usepackage{amssymb}
\usepackage{dsfont}
\usepackage{epsfig}
\usepackage{cite}

\newcommand{\comment}[1]{}

\newtheorem{thm}{Theorem}
\newtheorem{lem}[thm]{Lemma}
\newtheorem{prop}[thm]{Proposition}
\newtheorem{cor}[thm]{Corollary}

\newcommand{\reltxt}[2]{\stackrel{\text{\tiny #1}}{#2}}
\newcommand{\relas}[1]{\reltxt{a.s.}{#1}}

\DeclareMathOperator*{\essinf}{ess \mspace{3mu} inf}

\DeclareMathOperator*{\Prob}{Prob}
\DeclareMathOperator*{\ave}{ave}

\newcommand{\ind}{\mathds{1}}
\newcommand{\matsize}[2]{{#1}\mspace{-3mu}\times\mspace{-3mu}{#2}}

\newcommand{\Dmin}{D_{\min}}
\newcommand{\Dave}{D_{\ave}}

\renewcommand{\theenumi}{\roman{enumi}} 

\comment{

\newcommand{\captionfonts}{\small}

\makeatletter  
\long\def\@makecaption#1#2{%
  \vskip\abovecaptionskip
  \sbox\@tempboxa{{\captionfonts #1: #2}}%
  \ifdim \wd\@tempboxa >\hsize
    {\captionfonts #1: #2\par}
  \else
    \hbox to\hsize{\hfil\box\@tempboxa\hfil}%
  \fi
  \vskip\belowcaptionskip}
\makeatother   

}

%% file: genAEPjournalLong.bbl
\begin{thebibliography}{10}
\providecommand{\url}[1]{#1}
\csname url@rmstyle\endcsname
\providecommand{\newblock}{\relax}
\providecommand{\bibinfo}[2]{#2}
\providecommand\BIBentrySTDinterwordspacing{\spaceskip=0pt\relax}
\providecommand\BIBentryALTinterwordstretchfactor{4}
\providecommand\BIBentryALTinterwordspacing{\spaceskip=\fontdimen2\font plus
\BIBentryALTinterwordstretchfactor\fontdimen3\font minus
  \fontdimen4\font\relax}
\providecommand\BIBforeignlanguage[2]{{%
\expandafter\ifx\csname l@#1\endcsname\relax
\typeout{** WARNING: IEEEtran.bst: No hyphenation pattern has been}%
\typeout{** loaded for the language `#1'. Using the pattern for}%
\typeout{** the default language instead.}%
\else
\language=\csname l@#1\endcsname
\fi
#2}}

\bibitem{Ber:Random:1979}
H.~Berbee, \emph{Random Walks with Stationary Increments and Renewal Theory},
  ser. Mathematical Centre Tracts.\hskip 1em plus 0.5em minus 0.4em\relax
  Amsterdam: Mathematisch Centrum, 1979, vol. 112.

\bibitem{Chi:First:2001}
Z.~Chi, ``The first-order asymptotic of waiting times with distortion between
  stationary processes,'' \emph{IEEE Transactions on Information Theory},
  vol.~47, no.~1, pp. 338--347, Jan. 2001.

\bibitem{Chi:Stochastic:2001}
------, ``Stochastic sub-additivity approach to the conditional large deviation
  principle,'' \emph{Annals of Probability}, vol.~29, no.~3, pp. 1303--1328,
  2001.

\bibitem{DemKon:Asymptotics:1999}
A.~Dembo and I.~Kontoyiannis, ``The asymptotics of waiting times between
  stationary processes, allowing distortion,'' \emph{Annals of Applied
  Probability}, vol.~9, pp. 413--429, May 1999.

\bibitem{DemKon:Source:2002}
------, ``Source coding, large deviations, and approximate pattern matching,''
  \emph{IEEE Transactions on Information Theory}, vol.~48, no.~6, pp.
  1590--1615, June 2002.

\bibitem{DemZei:Large:1998}
A.~Dembo and O.~Zeitouni, \emph{Large Deviations Techniques and Applications},
  2nd~ed.\hskip 1em plus 0.5em minus 0.4em\relax New York: Springer, 1998.

\bibitem{Hol:Large:2000}
F.~den Hollander, \emph{Large Deviations}.\hskip 1em plus 0.5em minus
  0.4em\relax Providence: American Mathematical Society, 2000.

\bibitem{Har:First:2003}
M.~Harrison, ``The first order asymptotics of waiting times between stationary
  processes under nonstandard conditions,'' Brown University, Division of
  Applied Mathematics, Providence, RI, APPTS \#03-3, Apr. 2003.

\bibitem{HarKon:Maximum:2002}
M.~Harrison and I.~Kontoyiannis, ``Maximum likelihood estimation for lossy data
  compression,'' in \emph{Proceedings of the Fortieth Annual Allerton
  Conference on Communication, Control and Computing}, Allerton, IL, Oct. 2002,
  pp. 596--604.

\bibitem{Kal:Foundations:2002}
O.~Kallenberg, \emph{Foundations of Modern Probability}, 2nd~ed.\hskip 1em plus
  0.5em minus 0.4em\relax New York: Springer, 2002.

\bibitem{Kes:Sums:1975}
H.~Kesten, ``Sums of stationary sequences cannot grow slower than linearly,''
  \emph{Proceedings of the American Mathematical Society}, vol.~49, no.~1, pp.
  205--211, May 1975.

\bibitem{Kon:Implementable:1999}
I.~Kontoyiannis, ``An implementable lossy version of the lempel-ziv algorithm
  -- {P}art {I}: {O}ptimality for memoryless sources,'' \emph{IEEE Transactions
  on Information Theory}, vol.~45, no.~7, pp. 2293--2305, Nov. 1999.

\bibitem{LucSzp:Suboptimal:1997}
T.~{\L}uczak and W.~Szpankowski, ``A suboptimal lossy data compression based on
  approximate pattern matching,'' \emph{IEEE Transactions on Information
  Theory}, vol.~43, no.~5, pp. 1439--1451, Sept. 1997.

\bibitem{Roc:Convex:1970}
R.~T. Rockafellar, \emph{Convex Analysis}.\hskip 1em plus 0.5em minus
  0.4em\relax Princeton: Princeton University Press, 1970.

\bibitem{shannon:59}
C.~Shannon, ``Coding theorems for a discrete source with a fidelity
  criterion,'' \emph{IRE Nat. Conv. Rec.}, vol. part~4, pp. 142--163, 1959,
  reprinted in D. Slepian (ed.), {\em Key Papers in the Development of
  Information Theory}, IEEE Press, 1974.

\bibitem{Szp:Average:2001}
W.~Szpankowski, \emph{Average case analysis of algorithms on sequences}.\hskip
  1em plus 0.5em minus 0.4em\relax New York: John Wiley \& Sons, 2001.

\bibitem{yang-zhang:99c}
E.-H. Yang and Z.~Zhang, ``The shortest common superstring problem: {A}verage
  case analysis for both exact and approximate matching,'' \emph{IEEE Trans.
  Inform. Theory}, vol.~45, no.~6, pp. 1867--1886, 1999.

\bibitem{YanKie:Performance:1998}
E.-H. Yang and J.~C. Kieffer, ``On the performance of data compression
  algorithms based upon string matching,'' \emph{IEEE Transactions on
  Information Theory}, vol.~44, no.~1, pp. 47--65, Jan. 1998.

\bibitem{YanZha:Redundancy:1999}
E.-H. Yang and Z.~Zhang, ``On the redundancy of lossy source coding with
  abstract alphabets,'' \emph{IEEE Transactions on Information Theory},
  vol.~45, no.~4, pp. 1092--1110, May 1999.

\bibitem{ZhaYanWei:Redundancy:1997}
Z.~Zang, E.-H. Yang, and V.~K. Wei, ``The redundancy of source coding with a
  fidelity criterion -- {P}art~{I}: {K}nown statistics,'' \emph{IEEE
  Transactions on Information Theory}, vol.~43, no.~1, pp. 71--91, Jan. 1997.

\end{thebibliography}
